\documentclass{article}

\PassOptionsToPackage{numbers, compress}{natbib}

\usepackage[final]{neurips_2019}

\usepackage[utf8]{inputenc} 
\usepackage[T1]{fontenc}    
\usepackage{hyperref}       
\usepackage{url}            
\usepackage{booktabs}       
\usepackage{amsfonts}       
\usepackage{nicefrac}       
\usepackage{microtype}      
\usepackage{graphicx}

\title{Deep Learning for the Digital Pathologic Diagnosis of Cholangiocarcinoma and Hepatocellular Carcinoma: Evaluating the Impact of a Web-based Diagnostic Assistant}

\author{%
  Bora Uyumazturk\textsuperscript{1}\thanks{correspondence to: \texttt{buyumazturk@stanford.edu}}\hspace{1.5mm}\thanks{denotes equal primary  contribution}\hspace{1mm}, 
  Amirhossein Kiani\textsuperscript{1}\textsuperscript{\textdagger}, 
  Pranav Rajpurkar\textsuperscript{1}\textsuperscript{\textdagger}, 
  Alex Wang\textsuperscript{1}, 
  Robyn Ball\textsuperscript{2}, 
  \AND
  Rebecca Gao\textsuperscript{3}, 
  Yifan Yu\textsuperscript{1}, 
  Erik Jones, 
  Curtis P. Langlotz\textsuperscript{2, 5}, 
  Brock Martin\textsuperscript{4},
  Gerald J. Berry\textsuperscript{4},
  \And
  Michael G. Ozawa\textsuperscript{4},
  Florette K. Hazard\textsuperscript{4},
  Ryanne A. Brown\textsuperscript{4},
  Simon B. Chen\textsuperscript{4},
  Mona Wood\textsuperscript{4},
  \And
  Libby S. Allard\textsuperscript{4},
  Lourdes Ylagan\textsuperscript{4},
  Andrew Y. Ng\textsuperscript{1}\thanks{denotes equal senior contribution}\hspace{1mm},
  Jeanne Shen\textsuperscript{2, 4}\textsuperscript{\textdaggerdbl} \\
  \hspace{5mm} \\
  \textsuperscript{1}Department of Computer Science, Stanford University \\
  \textsuperscript{2}Center for Artificial Intelligence in Medical Imaging, Stanford University \\ 
  \textsuperscript{3}Stanford University School of Medicine\\ 
  \textsuperscript{4}Department of Pathology, Stanford University\\
  \textsuperscript{5}Department of Radiology, Stanford University
}

\begin{document}

\maketitle

\begin{abstract}
  While artificial intelligence (AI) algorithms continue to rival human performance on a variety of clinical tasks, the question of how best to incorporate these algorithms into clinical workflows remains relatively unexplored. We investigated how AI can affect pathologist performance on the task of differentiating between two subtypes of primary liver cancer, hepatocellular carcinoma (HCC) and cholangiocarcinoma (CC). We developed an AI diagnostic assistant using a deep learning model and evaluated its effect on the diagnostic performance of eleven pathologists with varying levels of expertise. Our deep learning model achieved an accuracy of 0.885 on an internal validation set of 26 slides and an accuracy of 0.842 on an independent test set of 80 slides. Despite having high accuracy on a hold out test set, the diagnostic assistant did not significantly improve performance across pathologists (p-value: 0.184, OR: 1.287 (95\% CI 0.886, 1.871)). Model correctness was observed to significantly bias the pathologist decisions. When the model was correct, assistance significantly improved accuracy across all pathologist experience levels and for all case difficulty levels (p-value: < 0.001, OR: 4.289 (95\% CI 2.360, 7.794)). When the model was incorrect, assistance significantly decreased accuracy across all 11 pathologists and for all case difficulty levels (p-value < 0.001, OR: 0.253 (95\% CI 0.126, 0.507)). Our results highlight the challenges of translating AI models to the clinical setting, especially for difficult subspecialty tasks such as tumor classification. In particular, they suggest that incorrect model predictions could strongly bias an expert's diagnosis, an important factor to consider when designing medical AI-assistance systems. 
\end{abstract}

\section{Introduction}
The rapid rate of scientific discovery in pathology has prompted a trend toward subspecialization \cite{conant_transition_2017, liu_trends_2016, sarewitz_subspecialization_2014} 
which has made it more difficult for general
surgical pathologists to confidently diagnose cases which lie outside of their areas of
expertise \cite{parkes_need_1997, rydholm_improving_1998}. Such situations are commonly
encountered during after-hours intraoperative
consultations, for example, when pathologists
are confronted with a diverse set of cases
across the range of specimen subtypes. A lack
of access to specialty expertise in this
context can result in slow clinical response
and adversely impact patient outcomes.

In this study, we explored the potential for AI to provide support to pathologists on subspecialty tasks by building and testing a deep learning-based assistant to help pathologists distinguish between hepatocellular carcinoma (HCC) and cholangiocarcinoma (CC), the two most common types of primary liver cancer. HCC and CC account for 70\% and 15\% of liver carcinomas, respectively, and each diagnosis has unique implications for biology, prognosis, and patient management \cite{amin_ajcc_2017, massarweh_epidemiology_2017}. For example, orthotopic liver transplantation is a widely accepted treatment for patients with HCC, but is often contraindicated in patients with CC. This diagnostic task is challenging, however, typically requiring the expertise of subspecialized gastrointenstinal (GI) pathologists \cite{altekruse_histological_2011, hass_subclassification_2018, lei_cytoplasmic_2006}. 

The assistant consisted of a cloud-deployed deep learning model and a browser-based interface where pathologists could receive a virtual second opinion on regions of interest in real time. The deep learning model was developed using a set of 20 publicly available hematoxylin and eosin (H\&E) slides, achieving an accuracy of 0.885 (95\% CI 0.710, 0.960) on the internal validation set. Using an independent test dataset, we evaluated the effect of assistance on the diagnostic accuracy, sensitivity, and specificity of eleven pathologists with varying levels of relevant experience. We did not find evidence that assistance improved overall pathologist accuracy (p-value: 0.184, OR: 1.287 (95\% CI 0.886, 1.871)). Upon further exploration, we found that pathologist accuracy was biased by model output, improving when the model was correct and worsening when the model was incorrect (p-value < 0.001). Our study highlights both the  challenges and potential benefits associated with deploying AI-based clinical support tools. 

\section{Methods}
\subsection{Data}

For model development, a total of 70 WSI (35 HCC and 35 CC) were randomly selected from the TCGA-LIHC and TCGA-CHOL  diagnostic slide collections from the Cancer Genome Atlas (TCGA). These were randomly partitioned into training, tuning and validation datasets. The training dataset (20 WSI) was used to learn model parameters, the tuning dataset (24 WSI) was used to choose hyperparameters, and the validation dataset (26 WSI) was used to assess the model’s generalizability to previously un-encountered data. 

The external validation set consisted of 80 representative WSI (40 CC and 40 HCC) of formalin-fixed, paraffin-embedded (FFPE) tumor tissue sections, one each from 80 unique patients randomly selected from a pool of all patients with HCC (250 patients) or CC (74 patients) who underwent surgical resection (partial or total hepatectomy) at a large academic medical center between the years 2011-2017, and who had glass slides available for retrieval from the pathology department archive \cite{grossman_toward_2016}. 

The reference standard diagnosis for all examinations was confirmed by re-review of all H\&E and immunostained slides from each case by a U.S. board-certified, GI/liver fellowship-trained subspecialty pathologist at a large academic medical center with 8 years of experience (J.S.). No diagnostic discrepancies were identified between the diagnosis rendered on the pathology report at the time of original interpretation and that of the reference pathologist.

\subsection{Model Development}

We trained a convolutional neural network to differentiate HCC from CC on image patches extracted from H\&E-stained digital WSI. We used the DenseNet-121 architecture, replacing the final layer to fit the binary prediction task \cite{huang_densenet}.

Because the entire WSIs were too large to input directly into the model, model training and evaluation were performed using smaller image patches extracted from the WSI. For each WSI in the training, tuning, and validation datasets, regions of interest (ROIs) containing tumor tissue were highlighted by the reference pathologist. For each slide, 1,000 square image patches of size 512 x 512 pixels were randomly sampled at 10x magnification from these ROIs. The pixel values for the patches were normalized using the mean and standard deviation of the pixel values from the TCGA training set before being used as input to the CNN. 

\subsection{Deep Learning Assistant}

\begin{figure}
  \includegraphics[width=11cm]{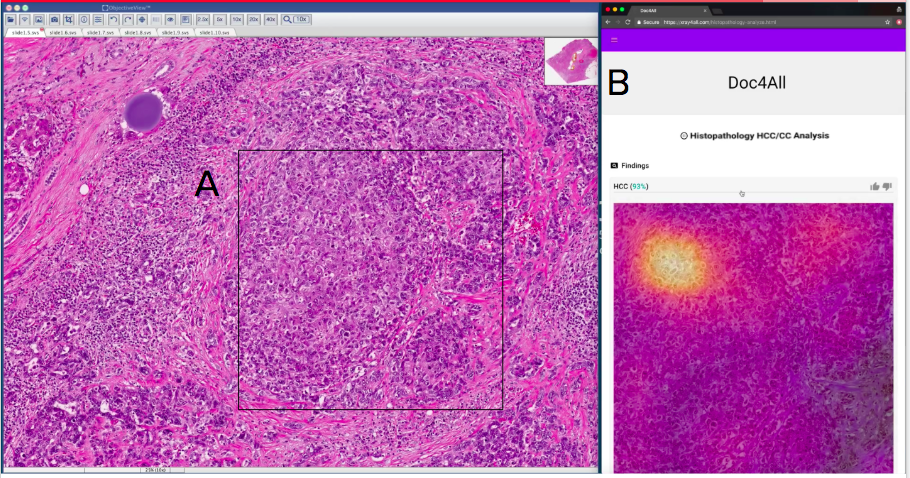}
  \centering
  \caption{Assistant Interface. When pathologists find a region of interest (A), they upload it to the web app (B) and receive the model class probabilities as well as a class-activation map.}
\end{figure}

In order to better assess the impact of our model in a clinical context, we created a browser-based interface through which pathologists could receive real-time feedback from the model. Pathologists would upload H\&E image patches and within seconds receive the model output. We chose to use patches as input so that pathologists could receive real-time feedback. The other alternative, running the model on the whole slide, patch-by-patch, would have been impractical given the context we were considering (after-hours intraoperative consultations) due to long processing times. To improve model interpretability, the model’s predictions were displayed as probabilities for each diagnosis (HCC or CC) and were accompanied by explanatory class activation maps (CAMs) which highlighted the image regions most consistent with each respective diagnosis (see Figure 1).

\subsection{Experiment Design}

We performed a diagnostic accuracy study comparing the performance of the pathologists with and without assistance. A total of eleven pathologists at a large academic medical center were recruited for the study. Of these, three were subspecialty GI pathologists who had spent the last three or more years diagnosing HCC and CC in daily independent practice, three were non-GI subspecialty pathologists with at least 12 years of independent practice experience who had not had routine diagnostic exposure to HCC or CC within the last five years, three were Anatomic Pathology residents (trainees) without independent diagnostic experience but with some exposure to HCC and CC cases as part of their training, and two  were not otherwise classified (NOC). 

To compare performance metrics of the pathologists with and without assistance, each pathologist participating in the study diagnosed the same set of 80 WSI in two separate sessions according to the crossover design detailed in Supplementary Figure 1. During each session they interpreted half of the study WSIs with assistance and half without. On assisted cases, they selected one or more patches from each WSI which they then uploaded to the assistant for real-time feedback. After viewing the outputs, they recorded their diagnosis. The pathologists were blinded to the original diagnoses, clinical histories, and follow-up information. After a washout period, the pathologists interpreted the same set of 80 WSI with the assistance status reversed (the WSI that were reviewed with assistance during the first reading were reviewed unassisted during the second reading, and vice versa). The pathologists were randomized into two groups, with one group beginning the diagnostic session with assistance, and the other without.

\subsection{Statistical Analysis}

We measured the accuracy of the model alone, the accuracies of the pathologists with and without assisted, and developed mixed effect models to test the significance of the impact of assistance on accuracy. We assessed the diagnostic accuracy of the model on the internal and external validation sets at the slide level by averaging patch level probabilities and converting this output to a binary slide-level prediction using a probability threshold of 0.5. 

The diagnostic accuracy of each group of pathologists, with and without assistance, was assessed on the external validation set. Average accuracy across all three subgroups was also computed. Confidence intervals were calculated using the Wilson Score method \cite{hartung_knapp}. 

Mixed-effect multivariate logistic regression models were developed to assess the impact of individual factors on pathologist accuracy. The random and fixed effects are enumerated in the Appendix. We evaluated the significance of assistance using a maximum likelihood method with the Wald Chi-square test. Using a similar method, we also evaluated the impact of correct and incorrect model outputs. Model development was conducted using the $\texttt{lme4}$ package in R \cite{bates_fitting_2015}.

\section{Results}

On the internal validation set, where the performance of the model was measured on all patches within the segmented tumor ROIs, the model achieved a diagnostic accuracy of 0.885 (95\% CI 0.710, 0.960). On the external validation set, where the performance of the model was measured on the patches input by the pathologists, the model alone achieved a mean accuracy of 0.842 (95\% CI 0.808, 0.876). 

There was not enough evidence to conclude that assistance improved pathologist performance (p-value: 0.184, OR: 1.287 (95\% CI 0.886, 1.871)). The accuracy of the eleven pathologists as a group was 0.89 (95\% CI 0.875, 0.916) without assistance, and 0.91 (95\% 0.893, 0.930) with assistance. The accuracy of trainees was 0.86 (0.95\% 0.809, 0.897) without assistance, and 0.90 (95\% 0.851, 0.928) with assistance. The accuracy of non-GI specialists was 0.84 (95\% CI 0.790, 0.882) without assistance and 0.87 (95\% CI 0.822, 0.910) with assistance. The accuracy of GI specialists was 0.95 (95\% 0.909, 0.968) without assistance and 0.96 (95\% CI 0.930, 0.980) with assistance. The accuracy of the NOC group was 0.97 (95\% CI 0.929, 0.987) without assistance and 0.93 (95\% CI 0.881, 0.961) with assistance. The performance of each group of pathologists with and without assistance is summarized in Supplementary Table 2 and presented in more detail in Supplementary Figure 3.

There is evidence that the model output biased pathologist diagnoses. Correct model predictions had a positive effect on accuracy (p-value: < 0.001, OR: 4.289 (95\% CI 2.360, 7.794)), while incorrect model predictions had a negative effect (p-value: < 0.001, OR: 0.253 (95\% CI 0.126, 0.507)).

\section{Conclusion}

In this study we explored the potential for AI to assist pathologists with a highly specialized diagnostic task. To do so, we built a deep learning assistant to assist with distinguishing between cancer subtypes in the liver and tested its impact on the diagnostic accuracy of pathologists with a range of experience levels. We found that overall pathologist accuracy did not improve using the model. Our results  raise questions about unintended effects of decision support tools. We found, for example, pathologist diagnoses were significantly biased both by correct and incorrect model outputs, suggesting a risk of anchoring. 

In conclusion, our study highlights the potential of AI to provide diagnostic support to physicians and presents methods for building and evaluating such a system. Before being considered for clinical deployment, however, further work must be done to test the robustness of such systems and measure the benefit that they provide. 
\newpage 

\small

\bibliography{references}

\newpage

\section*{Supplementary Material}

\renewcommand{\figurename}{Supplementary Figure}

\subsection{Model Training} 

The deep learning process consisted of feeding training images to the network, receiving a prediction from the network, and iteratively updating the parameters to decrease the prediction error, which was computed by comparing the network’s prediction to the known label for each image. By performing this procedure using a representative set of images, the resulting network could make predictions on previously unencountered H\&E-stained histopathology images. The weights of the network were initialized to those from a model pretrained on ImageNet, a large image classification dataset. The model was trained end-to-end, using stochastic gradient descent with a momentum of 0.9, on mini-batches of size 10. We used a step-based scheduler, which decayed the learning rate by a factor of 0.1 every 20,000 iterations. Learning rates were randomly sampled between 1e-4 and 1e-7.

We randomly extracted 1,000 image patches of size 512 x 512 pixels from annotated tumor regions within each whole-slide image (WSI) in the training set, resulting in a total of 20,000 training patches, which served as training inputs into the model. From the tuning and internal validation datasets, we randomly extracted 100 similarly sized image patches per WSI, for a total of 2,400 tuning patches and 2,600 validation patches. Pixel values were normalized to the means and standard deviations of images in the training set. To improve the generalizability of the models, several forms of data augmentation were used during training, including rotations and flips of the input images. 

\subsection{Model Selection}

Model selection consisted of three steps. First, 50 networks with randomly sampled hyperparameters were trained on the TCGA training dataset, and evaluated on the tuning set. From these, the 10 best-performing networks were selected and evaluated on the internal validation set, to assess generalizability to unencountered data. The network with the highest accuracy on the internal validation set was used to create the assistant. 

\subsection{Assistant Web Architecture}
The assistant’s web architecture is comprised of an HTML5 front end and a Python back end (see Supplementary Figure 3). The front end communicates with the back end via a JSON-based REST interface. The front end is responsible for authenticating the users, allowing users to upload patches, viewing the model results and exploratory CAMs in real time, and providing feedback about the model’s output. 

\subsection{Model Explanations}

Class activation maps (CAMs) were used to highlight regions with the greatest influence on the model’s decision. For a given patch, the CAM was computed for both classes (HCC and CC) by taking the weighted average across the final convolutional feature map, with weights determined by the linear layer. The CAM was then scaled according to the output probability, so that more confident predictions appeared brighter. Finally, the map was upsampled to the input image resolution, and overlaid onto the input image.

\subsection{Mixed Effect Modeling} Mixed-effect multivariate logistic regression models were developed to assess the impact of individual factors on pathologist accuracy. Assistance status (assisted vs. unassisted), pathologist subgroup (GI specialist, non-GI specialist, or trainee), ground truth, experiment order (whether the pathologist had assistance first or not), and tumor grade (well-differentiated, moderately differentiated, or poorly differentiated, corresponding to grades 1, 2,  and 3) were included as fixed effects, while the pathologist and slide were included as random effects.
\newpage

\subsection{Supplementary Figures and Tables}

\setcounter{figure}{0}
\begin{figure}[h]
    \centering
    \includegraphics[width=12cm]{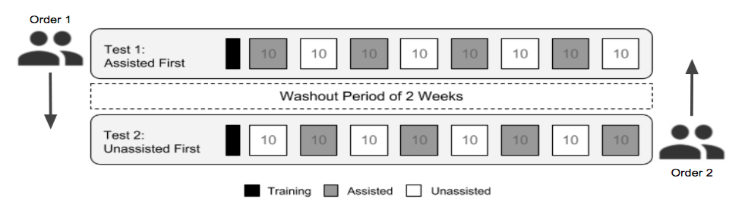}
    \caption{Each of the 11 pathologists was randomly assigned to either test order 1 or 2. Each test began with a brief training block of 4 (2 HCC and 2 CC) practice whole-slide images (WSI), followed by 8 experiment blocks of 10 WSI each, with order 1 beginning with assistance, and order 2 beginning without assistance. The same 80 experiment WSI were reviewed in the same sequence during Tests 1 and 2, across both test orders.}
    \label{fig:my_label}
\end{figure}

\begin{figure}[h]
    \centering
    \includegraphics[width=10cm]{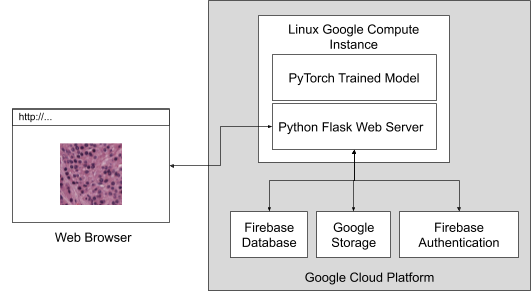}
    \caption{The assistant’s web architecture is comprised of an HTML5 front end and a Python back end. The front end communicates with the back end via a JSON-based REST interface. The front end is responsible for authenticating the users, allowing users to upload patches, viewing the}
    \label{fig:my_label}
\end{figure}

\begin{figure}[h]
    \centering
    \includegraphics[width=12cm]{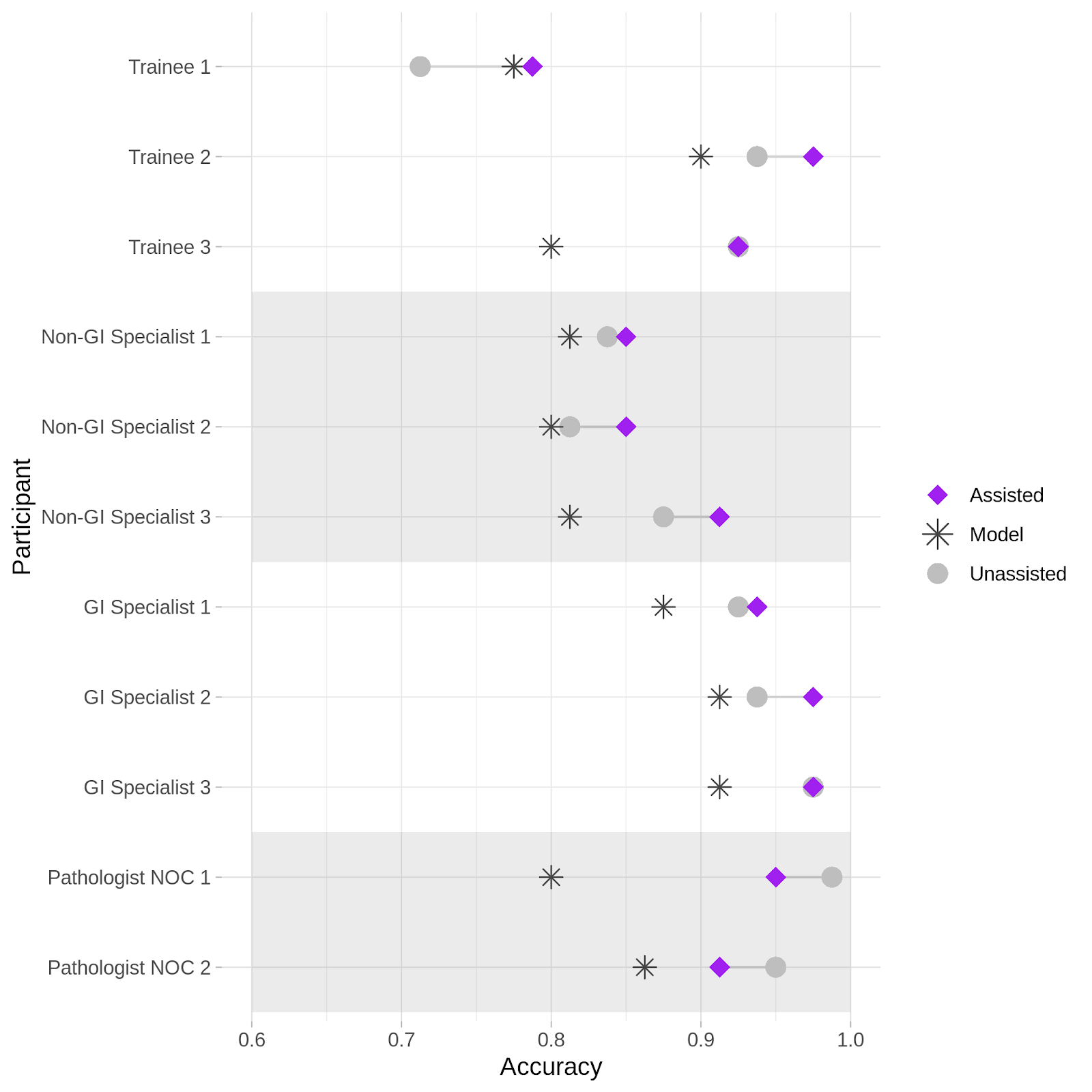}
    \caption{The average diagnostic accuracy (across the set of 80 experiment WSI) for each pathologist is plotted as follows: gray circle (Unassisted) = accuracy of the unassisted pathologist, star (Model) = accuracy of the model alone (based on pathologist selected input patches), purple diamond (Assisted) = accuracy of the pathologist with model assistance.}
    \label{fig:my_label}
\end{figure}

\renewcommand{\tablename}{Supplementary Table}

\begin{table}[h]
  \caption{Pathologist unassisted and assisted accuracies}
  \label{sample-table}
  \centering
  \begin{tabular}{lll}
    \toprule
    Experience Level    & Unassisted (95\% CI) & Assisted (95\% CI) \\
    \midrule
    Trainee & 0.858 (0.809, 0.897)  & 0.896 (0.851, 0.928)     \\
    Non-GI Specialist     & 0.842 (0.790, 0.882) & 0.870 (0.822, 0.910)     \\
    GI Specialist     & 0.946 (0.909, 0.968)       & 0.962 (0.930, 0.980)  \\ 
    NOC & 
    0.969 (0.929, 0.987) & 0.931 (0.881, 0.961) \\ 
    \midrule
    All Pathologists & 0.897 (0.875, 0.916) & 0.913 (0.893, 0.930) \\
    \bottomrule
  \end{tabular}
\end{table}

\end{document}